# Conductance versus current noise in a neuronal model for noisy subthreshold oscillations and related spike generation

Martin Tobias Huber[1] And Hans Albert Braun[2]

[1]*Department of Psychiatry and Psychotherapy, University of Marburg, Rudolf-Bultmannstraße 8, D-35033 Marburg, Germany* and [2]*Institute of Normal and Pathological Physiology, University of Marburg, Deutschhausstraße 1, D-35033 Marburg, Germany*

**ABSTRACT**

Biological systems are notoriously noisy. Noise, therefore, also plays an important role in many models of neural impulse generation. Noise is not only introduced for more realistic simulations but also to account for cooperative effects between noisy and nonlinear dynamics. Often, this is achieved by a simple noise term in the membrane equation (current noise). However, there are ongoing discussions whether such current noise is justified or whether rather conductance noise should be introduced because it is closer to the natural origin of noise. Therefore, we have compared the effects of current and conductance noise in a neuronal model for subthreshold oscillations and action potential generation. We did not see any significant differences in the model behavior with respect to voltage traces, tuning curves of interspike-intervals, interval distributions or frequency responses when the noise strength is adjusted. These findings indicate that simple current noise can give reasonable results in neuronal simulations with regard to physiological relevant noise effects.

# 1 INTRODUCTION

Biological systems are notoriously noisy. Noise, therefore, also plays an important role in many models of neural impulse generation and the noise is not only introduced for more realistic simulations but also to account for cooperative effects between noisy and nonlinear dynamics. One interesting example are subthreshold membrane potential oscillations and associated action potential generation in various neurons in the central and peripheral nervous system (Pare et al. 1995, Braun et al. 1994, Xing et al. 2001 and many others). Naturally present stochastic fluctuations play an important role for the response behavior in the situation when oscillations are close to the spike threshold and where the noise essentially determines spike initiation (Braun et al.1980, 1994, 2003, White et al. 1998). In addition mechanisms such as coherence and/or stochastic resonance can occur which also might contribute to the response properties of the neurons (Gammaitoni et al. 1998).

Different sources of stochastic fluctuations thereby will contribute to and influence the ionic dynamics. Membrane noise includes fluctuations in the ionic conductances including the influence of stochastic synaptic background activity (e.g. from ongoing activity in cortical networks), thermal current noise, noise due to electrogenic ion pumps as well as fluctuations from changes in the environment of the neuron (Tuckwell 1988). The sources of stochastic fluctuations will vary depending on the actual location of the neuron, e. g. whether it is located in the central or peripheral nervous system and in particular whether it is subjected to synaptic input or not. The latter is the case for primary peripheral sensory receptors which represent free nerve endings and hence do not receive any synaptic input. However, also in such receptors noise is known to play an important role for signal encoding.

In contrast, balanced random synaptic activity is a major noise source in central nervous system neurons and the role of the synaptic background activity for stimulus-response relations and in particular neuronal gain control has been emphasized in recent studies (Ho and Destexhe 2000, Chance et al. 2002 and others). However, the effects of the different noise sources cannot be seen in isolation because noise sources not necessarily act independently and the strength of one noise source might change on changes of other state variables (a comprehensive discussion is given by Longtin and Hinzer (1995).

Different approaches have been used to account for the role of stochastic fluctuations in neuronal models and representative examples for different levels of complexity are the following: Ho and Destexhe (2000) use a sophisticated physiologically-detailed approach and model random synaptic release conditions in a way that synaptic conductance and correlation parameters can be varied independently. White et al. (1998) introduce noise into a model for subthreshold oscillations by modelling stochastic persistent sodium channels as a two-state Markov process. Longtin and Hinzer (1996) in contrast simulate membrane noise by a current noise term in the membrane equation. The noise term there includes pump noise, thermal noise and effects of conductance noise and is modelled as gaussian distributed exponentially correlated noise. In our own modelling studies on signal encoding with neuronal oscillations (e.g. Braun et al. 1998, 2003) we simply used gaussian white noise and this approach is also often used in theoretical neuronal studies (e.g. Lindner et al. 1999, 2002).

The question arises, whether results from simulation studies using one or another implementation of noise sources hold true for the different cases. We were interested here, whether the stimulus-response properties of our ionic subthreshold oscillator model (Huber and Braun 2005) obtained with the current noise term in the membrane equation are comparable to those obtained with a different implementation of noise, that is for example with stochastic fluctuations of a specific ionic conductance (conductance noise). This is relevant as we discuss the stimulus-response and neuromodulatory properties of subthreshold oscillations and noise for different levels of the nervous system, i.e. specific cells and functions such as nociception and dorsal root ganglion cells or neuromodulation and layer II entorhinal cortex neurons.

In this brief study we compare simulations with current noise with simulations which were obtained by introducing white noise into the model´s potassium conductance $g_K$. Model and noise implementation are described in section 2. In section 3 we demonstrate simulated voltage time traces, plots of successive interval durations, interspike interval histograms as well as the mean spike frequency versus applied current $I_{app}$, respectively, for the two cases: current noise and $g_K$-conductance noise. We end with concluding remarks in section 4.

## 2 MODEL

The model represents an ionic conductance model for subthreshold oscillations and action potential generation (Huber and Braun 2005). The model consists of two sets of simplified sodium and potassium conductances operating at two different levels of the membrane potentials and two different time scales: a Hodgkin/Huxley-type spike encoder is represented by rapid, high-voltage activating $g_{Na}$ and $g_K$, whereas the subthreshold oscillations essentially depend on the interplay of a persistent sodium conductance, $g_{Nap}$, and a subthreshold potassium conductance, $g_{Ks}$

$$C_M \, dV/dt = - I_l - I_{Nap} - I_{Ks} - I_{Na} - I_K + I_{app} \tag{1}$$

with $C_M = 1\mu F/cm^2$ the membrane capacity, V the membrane voltage, $I_l = g_l (V-V_l)$ a leak current with $g_l = 0.1$ mS/cm$^2$ and $V_l = -60$ mV. $I_{app}$ is injected current. The voltage-dependent currents $I_{Nap}$, $I_{Ks}$, $I_{Na}$ and $I_K$ are modelled as

$$I_i = g_i \, a_i \, (V - V_i) \tag{2}$$

with $g_i$ the respective maximum conductances (i denotes Na, Nap, K, Ks), $a_i$ the voltage-dependent activation variables and $V_i$ the respective Nernst potentials. The activation variables are given as

$$\tau_i \, da_i / dt = F_i - a_i \tag{3}$$

with

$$F_i = 1/\{1 + \exp[-s_i (V - V_{0,i})]\} \tag{4}$$

where the $\tau_i$ are time constants, $s_i$ the slope and $V_{0,i}$ the half-activation potentials. Activation of $I_{Na}$ is instantaneous, thus $a_{Na} = a_{Na\infty}$.

The noise $\zeta$ is added to the model in two different ways: either as current noise or as conductance noise. In the case of current noise, $\zeta$ is an additional term in the membrane equation which then is given as

$$C_M \, dV/dt = - I_l - I_{Nap} - I_{Ks} - I_{Na} - I_K + I_{app} + \zeta \tag{1a}$$

and in the case of conductance noise we introduce the noise term into the differential equation for the activation variable $a_K$ of the potassium conductance $g_K$:

$$\tau_K \, da_K / dt = F_K - a_K + \zeta \tag{5}$$

In both cases $\zeta$ is gaussian white noise with the properties $\langle \zeta(t) = 0 \rangle$ and $\langle \zeta(t)\zeta(s) \rangle = 2D\,\delta(t-s)$ which determines all of its statistical features. In the case of current noise equation (1) is replaced by equation (1a). In contrast, in the case of conductance noise the equation for the activation variable of the $g_K$ (eq. 3 with i = K) is replaced by equation (5).

The system of equations was solved numerically by use of the forward Euler integration method with stepsize adjusted to 0.1 ms according to the implementation of Fox et al. (1988). Numerical parameter values: $V_{Na} = 50$, $V_K = -90$, $g_{Na} = 2.0$, $g_K = 2.0$, $g_{Nap} = 0.4$, $g_{Ks} = 2.0$, $s_{Na} = s_K = s_{Nap} = s_{Ks} = 0.25$, $\tau_K = 2.0$, $\tau_{Nap} = 10$, $\tau_{Ks} = 50$, $V_{0Na} = V_{0K} = -25$, $V_{Nap} = V_{0Ks} = -40$. Systems of units is ms, mV, mS/cm$^2$, mA/cm$^2$.

## 3 RESULTS

In the following we consider the responses of the model to application of depolarising current and with respect to the two different noise situations – current noise and $g_K$-conductance noise (figure 1a,c). On depolarising $I_{app}$ and in the presence of noise, the model exhibits subthreshold oscillations in membrane potential which rise in amplitude when the current strength increases. Once oscillations are close to the spike threshold, the stochastic fluctuations become important for the generation of action potentials. With further increasing $I_{app}$, periodic spiking results because oscillations cross the spike threshold (almost) each cycle. The voltage traces demonstrated in figure 1a (current noise, D = 0.1) and figure 1c ($g_K$-conductance noise, $D = 2*10^{-5}$) thereby are remarkably similar. No significant differences can be seen. The only real difference is the absolute value of the noise strength for the two respective cases.

This finding is confirmed by the distribution of interspike interval durations in response to a continuous change of the applied current (figure 1b,c, upper graphs; ramp-shaped change in $I_{app}$ from 1.0 -> 2.0 mA/cm$^2$) with respect to the two different noise sources. In addition, when we look at the spike statistics demonstrated by interspike interval histograms (ISIHs) obtained from long simulation runs (figure 1b,d, lower graphs) we cannot observe differences in the distribution and heights of the interval peaks. Importantly, the multimodality of the ISIH indicative for mixed patterns of subthreshold oscillations and action potentials is well preserved also in the case of $g_K$-conductance noise.

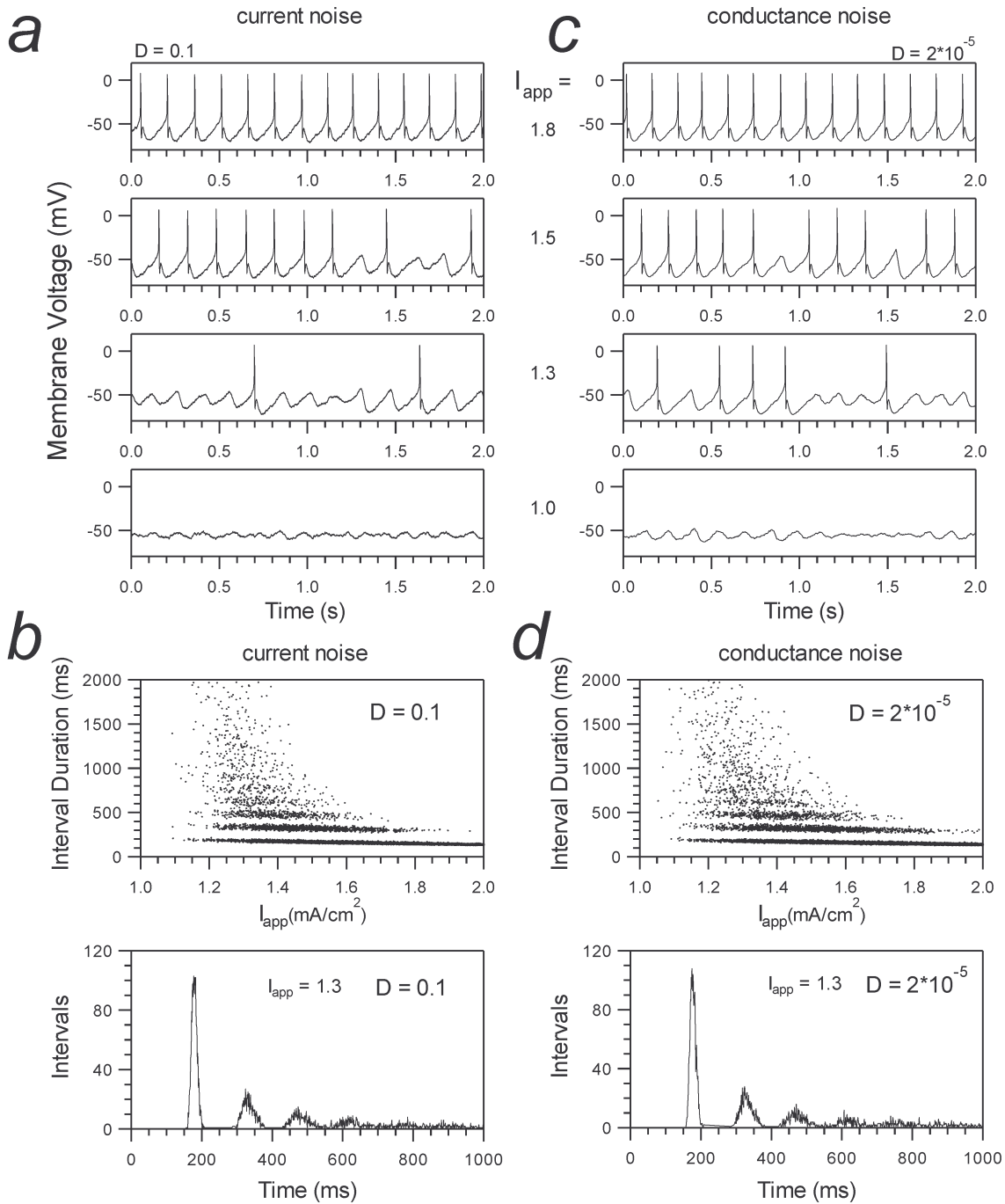

**Figure 1:** Responses of the model with respect to current noise (graphs *a* and *b* on the left side of the figure) versus conductance ($g_K$) noise (graphs *c* and *d* on the right side of the figure). (*a*) and (*c*) voltage traces obtained at different levels of depolarizing current $I_{app}$ (1.0, 1.3, 1.5 and 1.8 mA/cm$^2$). The upper graphs in (*b*) and (*d*) show time plots of successive interspike intervals on response to a ramp-shaped change of the applied current (time = 5000 s, increment $\Delta I_{app}$ = 0.0002 mAcm$^{-2}$ms$^{-1}$). The lower graphs in (*b*) and (*d*) show interspike interval histograms obtained from simulations with $I_{app}$ = 1.3 mA/cm$^2$ (each plot contains N = 5000 intervals). The noise strength is D = 0.1 in the case of current noise and D = 2*10$^{-5}$ in the case of conductance noise.

Finally, we calculated the mean spike frequencies F in dependence of the applied current $I_{app}$. Figure 2 demonstrates the $F - I_{app}$ curves for current noise (dashed curves) and $g_K$-conductance noise (solid curves). Curves are obtained from simulations with two different noise strengths for each noise case: D = 0.1 and 1.0 for current noise and D = $2*10^{-5}$ and D = $2*10^{-4}$ for $g_K$-conductance noise. The $F - I_{app}$ curves show that in both noise cases equal stimulus-response relations are obtained for the range of $I_{app}$ values chosen. Also, changing the noise strength in both cases by a factor of 10 does equally change the $F - I_{app}$ curves in both cases, that is, to more linearized stimulus-response curves. The $F - I_{app}$ curves resulting from the higher noise strength thereby again are almost identical.

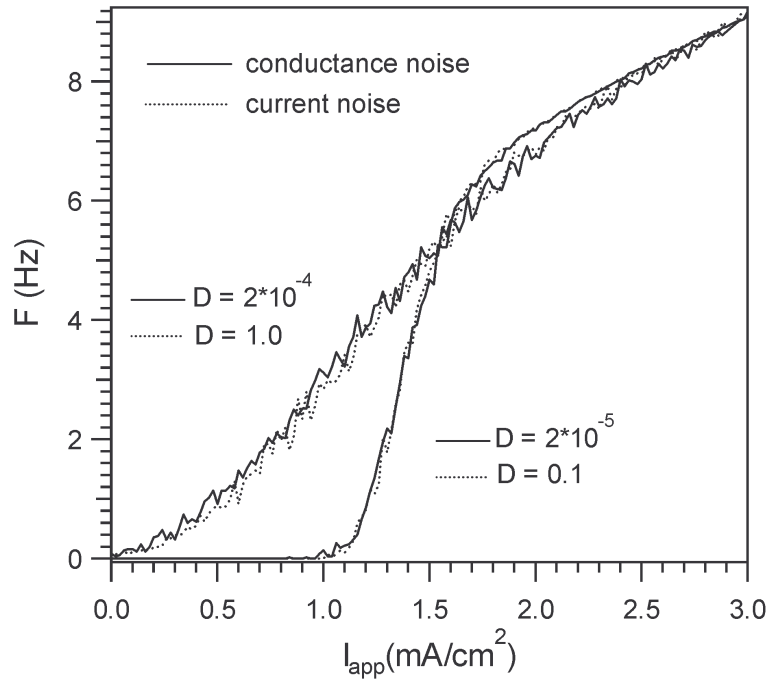

**Figure 2:** Mean spike frequency F (Hz) versus $I_{app}$ (mA/cm$^2$) for the two different noise situations - current noise (dashed lines) and conductance noise (solid lines) and for two different values of the noise strength (current noise: D = 0.1 and 1.0; conductance noise D = $2*10^{-5}$ and D = $2*10^{-4}$.

## 4 CONCLUSION

In summary we compared the influence of a current noise term with the influence of a conductance noise term on the responses of a previously described ionic neuronal model for subthreshold membrane potential oscillations and related spike generation. Such types of neurons are found in various regions of the peripheral and central nervous system and therefore underlie different sources of membrane noise as well as serving different functional purposes. On the other hand, realistic descriptions of noise are a complex issue and approaches at very different levels of complexity are used in neuronal modelling studies (e.g. Ho and Destexhe 2000, White et al. 1998, Longtin and Hinzer 1996, Braun et al. 2003, Lindner et al. 1999, 2002). One relevant question thereby is, whether a simple gaussian white noise term in the membrane equation, as often used, is justified at least in some instances or whether different noise sources, in particular conductance noise, have to be modelled explicitly.

In our brief orientating study here, we examined the effects of the two different noise sources on voltage traces, plots of successive interspike interval durations, interspike interval histograms and $F - I_{app}$ curves. Apart from the necessarily different absolute values of the noise strengths in the two respective cases, we found no significant differences in the response behaviours of our specific model. Although our study does not represent a comprehensive account and also very different conductance noise implementations could be considered (and indeed might be important in physiological reality), the findings nevertheless indicate that in our case the specific implementation of the noise is not of critical importance. This is also indicated by orientating simulations where we added the noise not to the $g_K$ activation variable but to the maximum conductances and where by adjusting of the noise strength we got comparable simulation results. In other words, for the type of signal encoding and neuromodulation we are interested in with our model, the important issue seems to be the presence of some kind of stochastic fluctuations but the actual implementation of the fluctuations as current or conductance noise is not of critical importance. The noise-mediated interactions of oscillations and spike generation mechanisms will work as long as appropriate voltage fluctuations occur. With this respect the model also behaves robust to tuning of the noise strength. Bifurcation plots and voltage traces as shown in the simulations here or in Huber and Braun (2005) are not restricted to a narrow range of noise strength.

However, differences can become critical when the actual source of noise and when the actual ingredients of the noise have to be considered for a given neuronal behaviour. For example, investigation of the effects of conductance and coherence properties of random synaptic background activity would necessitate physiologically more detailed modelling approaches. In a next step one than might be able to reduce the complexity in accordance with the biological behavior and, for example, injection of noisy current in a model might represent voltage fluctuations which to some extend are comparable to those due to random synaptic background activity. A critical assessment will be necessary as such simplifications might be correct in one situation but could lead to false results with other problems. In addition, from an information processing perspective, it will also be important to consider whether the noise represents an external independent source or whether it is inside the transmission system. In the latter case it might be oversimplifying or even false to treat noise separate from the signal as the noise might change with the signal (see Greenwood and Lansky 2005 for thorough discussion). Nevertheless, our preliminary study altogether indicates that gaussian white noise in the membrane equation seems to be a useful first approximation to stochastic neuronal activity under defined conditions.

**Acknowledgment:** We like to thank Lutz Schimansky-Geier for most valuable discussions on noise and nonlinear dynamics. We acknowledge financial support by the EU Network of Excellence NoE 005137 BioSim (RA5WP10).